\documentclass[twocolumn,twoside,slac_two]{revtex4}
\usepackage{graphicx}

\usepackage{amsmath}  
\usepackage[binary-units]{siunitx}
 
\newcommand{\virgolette}[1]{\lq\lq #1\rq\rq}

\usepackage{fancyhdr}
\usepackage{hyperref}
\hypersetup{
    colorlinks=true,
    urlcolor=magenta,
}
\begin{document}

\title{The electronics of the HEPD of the CSES experiment }

\author{V.~Scotti, ~
        G.~Osteria}
\affiliation{INFN Napoli,  Complesso Universitario di M. S. Angelo, Ed. 6- Via Cintia, 80126 Napoli, Italy}
       \author{for the CSES-Limadou Collaboration}
\affiliation{http://cses.roma2.infn.it/ }

\begin{abstract}
The China Seismo Electromagnetic Satellite (CSES) aims to contribute to the monitoring of earthquakes from space. This space mission, lead by a Chinese-Italian collaboration, will study phenomena of electromagnetic nature and their correlation with the geophysical activity. The satellite will be launched in 2017 and will host several instruments onboard: two magnetometers, an electrical field detector, a plasma analyzer, a Langmiur probe and the High Energy Particle Detector (HEPD). The HEPD, built by the Italian collaboration, will study the temporal stability of the inner Van Allen radiation belts, investigating precipitation of trapped particles induced by magnetospheric, ionosferic and tropospheric electromagnetic emissions, as well as by seismo-electromagnetic disturbances. It consists of two layers of plastic scintillators for trigger and a calorimeter. The direction of the incident particle is provided by two planes of double-side silicon microstrip detectors. HEPD is capable of separating electrons and protons and identify nuclei up to Iron. The HEPD will study the low energy component of cosmic rays too. The HEPD comprises the following subsystems: detector, electronics, power supply and mechanics. The electronics can be divided in three blocks: silicon detector, scintillator detectors (trigger, energy and veto detectors) and global control and data managing. In this paper a description of the electronics of the HEPD and its main characteristics will be presented.
\end{abstract}

\maketitle

\thispagestyle{fancy}


\section{Introduction}

CSES is a scientific mission dedicated to study electromagnetic, plasma and particles perturbations of atmosphere, ionosphere, magnetosphere and Van Allen belts induced by natural sources and anthropocentric emitters and their correlations with the occurrence of seismic events. 
 
Among the possible phenomena generated by an earthquake, bursts of Van Allen belt electron fluxes in the magnetosphere have been repeatedly reported in literature by various experiments, though a statistical significance was always difficult to claim \cite{Wang,Zhang,Sgrigna}. The CSES mission aims at measuring such particle bursts, by means of the High Energy Particle Detector (HEPD). The high inclination orbit of the satellite allows the instrument to detect particles of different nature during its revolution: galactic cosmic rays - which are modulated by the solar activity at low energies and also solar energetic particles associated to transient phenomena such as Solar Flares or Coronal Mass Ejections.

The satellite will be placed at a \SI{98}{\degree} Sun-synchronous circular orbit at an altitude about 500 km, the launch is scheduled in July 2017 with an expected lifetime of 5 years. The satellite mass will be about 730 kg and the peak power consumption about 900 W. 

CSES satellite hosts several instruments on board (see Figure \ref{fig:instrument}):
\begin{figure}[htb] 
\centering 
\includegraphics[width=0.9\linewidth]{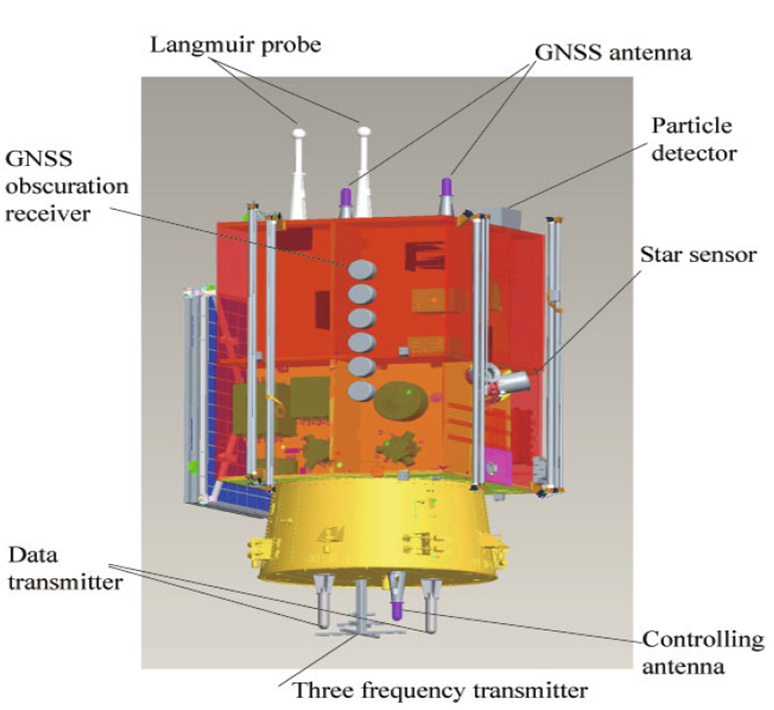}
\caption{Instruments onboard the CSES satellite. }
\label{fig:instrument}. 
\end{figure}
\begin{itemize}
\item a Search-Coil Magnetometer, a High Precision Magnetometer and an Electric Field Detector for measuring the magnetic and electric fields;
\item a Plasma Analyser Package and a Langmuir Probe for measurements of local plasma disturbances;
\item a GNSS Occultation Receiver and a Transmitter for the study of profile disturbance of plasma;
\item the High-Energy Particle Package and High-Energy Particle Detector (HEPD) for the measurement of the flux of energetic particles.
\end{itemize}

CSES is the first satellite of a space monitoring system proposed in order to investigate the topside ionosphere - with the most advanced techniques and equipment - and designed in order to gather world-wide data of the near-Earth electromagnetic environment.

\section{The High Energy Particle Detector}

The High-Energy Particle Detector has been developed by the Italian CSES collaboration, due to its long experience in cosmic ray physics. CSES will complement the cosmic ray measurements of PAMELA \cite{pamela} at low energy, thus giving a complete picture of the cosmic ray radiation by direct measurements from the very low energies (few MeVs) up to the the TeV region.

The High Energy Particle Detector (HEPD) will study low energy Cosmic Rays (CR) in the energy range 3 - 300 MeV. The HEPD has to separate electrons and proton, identifying electrons within a proton background (N$_e$/N$_p=$ \SI{e-5}{}$\div$\SI{e-3}{}), and identify nuclei up to Iron. The high-inclination orbit allows the telescope to detect particles of different nature during its revolution: galactic CR, Solar Energetic Particles, particles trapped in the magnetosphere.

The HEPD comprises the following subsystems: detector, electronics, power supply and mechanics. The detector is contained in an aluminum box, while the electronics cards are placed outside the detector fixed on the base plate by means of a dedicated supporting structure. The outside surface is covered with aluminized polyimide layer to assure a good thermal insulation.

\begin{figure}[htb] 
\centering 
\includegraphics[width=\linewidth]{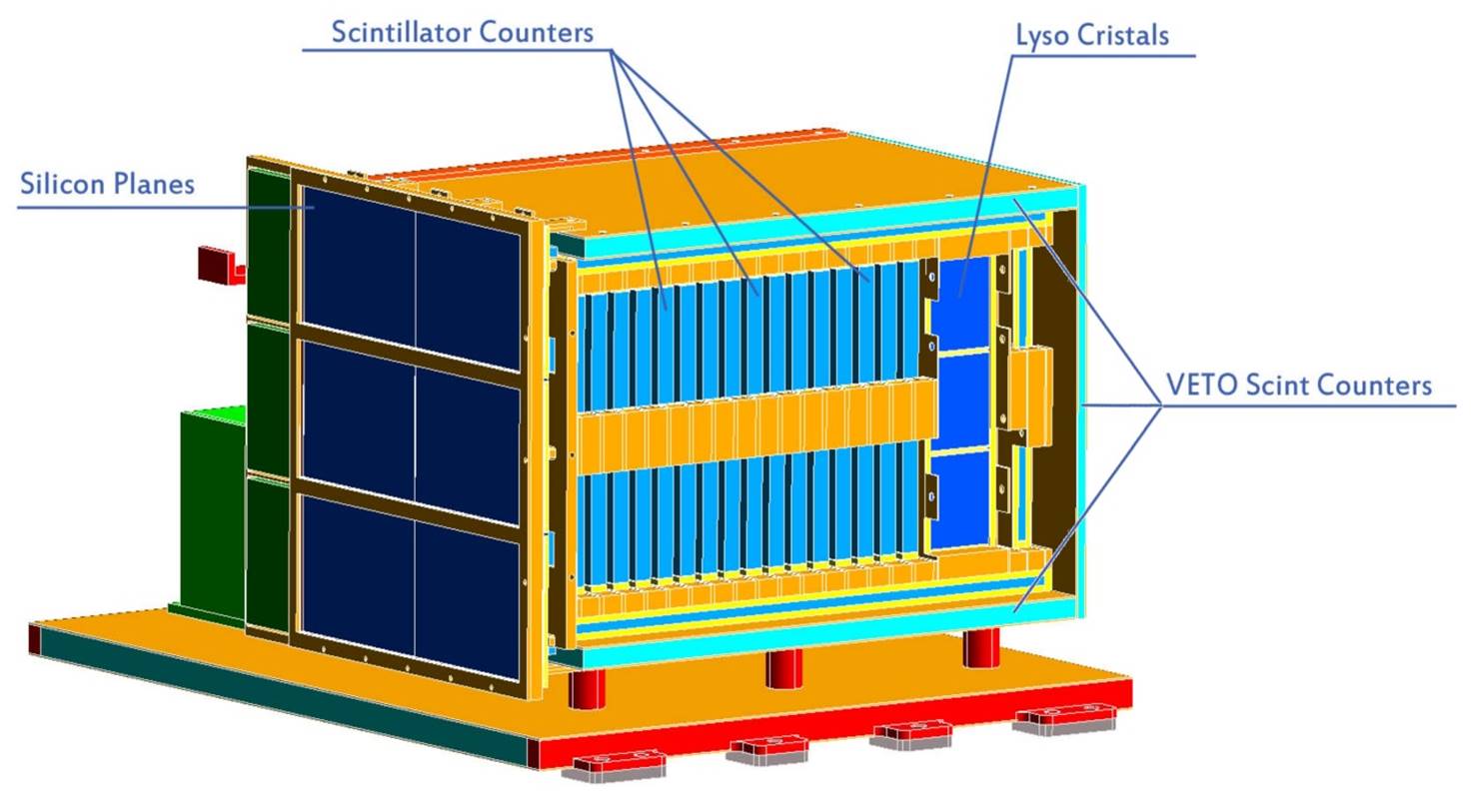}
\caption{An overview of the HEPD instrument. }
\label{fig:hepd}. 
\end{figure}

The detector consists of three components:
\begin{itemize}
\item Silicon planes: two planes of double-side silicon micro-strip detectors are placed on the top of the detector in order to track the direction of the incident particle limiting the effect of Coulomb multiple scattering on the direction measurement;
\item Trigger: a layer of thin plastic scintillator divided into six segments;
\item Calorimeter: a tower of 16 layers of \SI{1}{\centi\meter} thick plastic scintillator planes followed by a 3$\times$3 matrix of an inorganic scintillator LYSO.
\end{itemize}
An organic scintillator is used in the calorimeter to optimize the energy resolution. In order to extend the electron measurement range to 100 MeV an inorganic scintillator LYSO is used for the last plane of the calorimeter. The calorimeter volume is surrounded by \SI{5}{\milli\meter} thick plastic scintillator veto planes. All the scintillator detectors (trigger, calorimeter and VETO) are read out by photomultiplier tubes (PMTs).

The good energy-loss measurement of the silicon track, combined with the energy resolution of the scintillators and calorimeter, allows identifying electrons with acceptable proton background levels (N$_e$/N$_p=$ \SI{e-5}{}$\div$\SI{e-3}{}).
In table \ref{tab:tab1} the main parameter of the detector are summarized. 

\begin{table}
 \centering
  \begin{tabular}{cc}
    \hline
    Parameter & Value\\
    \hline
    Energy Range & Electrons: 3-100 MeV \\
     & Protons: 30-300 MeV \\
    Angular resolution & $<8^{\circ}$ 5 MeV \\
    Energy resolution & $<10\%$ 5 MeV \\
    Particle identification & $>90\%$ \\
    Free field of view & $\geq 70^{\circ}$ \\
    Pointing & Zenith \\
    Operative temperature & -10$^{\circ}$+45$^{\circ}$ \\
    Mass & $<$\SI{35}{\kilo\gram} \\
    Power Consumption & $<$\SI{38}{\watt} \\
    Mechanical dimensions &  20$\times$20$\times$\SI{40}{\centi\meter\cubed}\\
     \hline
  \end{tabular}
  \caption{HEPD main technical characteristics.}
  \label{tab:tab1}
\end{table}

The Italian collaboration developed four models of the HEPD: Electrical, Mechanical and Thermal, Qualification (QM) and Flight Models.

\section{The electronics of the HEPD}

The electronics of the HEPD can be divided in three blocks: 
\begin{itemize}
\item Silicon detector;
\item Scintillator detectors;
\item Global control and data managing. 
\end{itemize}

\begin{figure}[htb] 
\centering 
\includegraphics[width=0.9\linewidth]{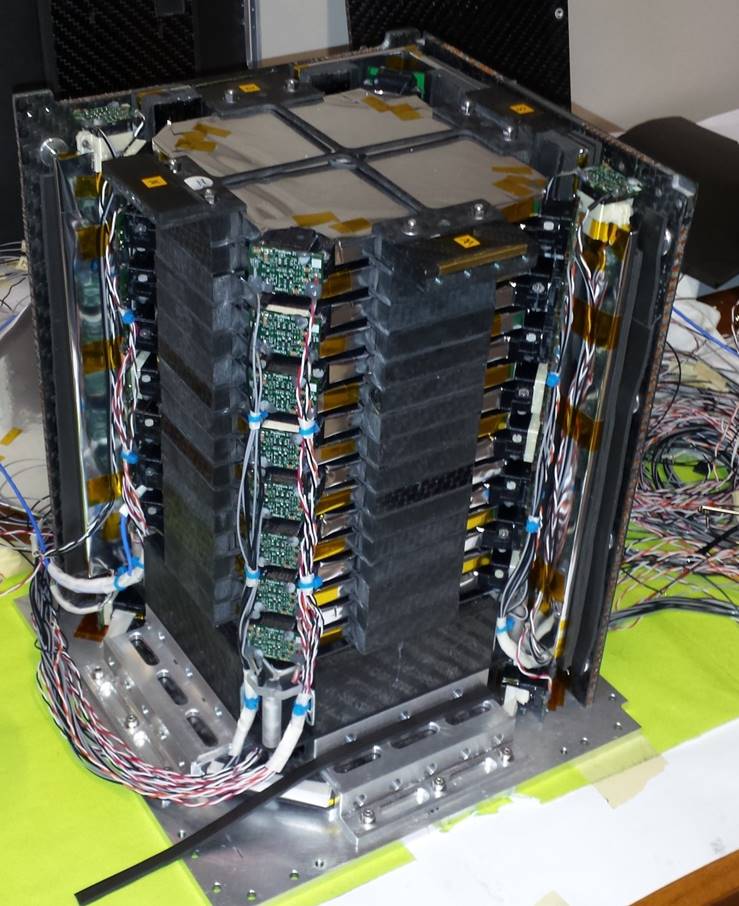}
\caption{Side view of the QM calorimeter which shows the plastic scintillator planes. The PMTs are at the corners of each calorimeter plane.
 }
\label{fig:calo}. 
\end{figure}
Each detector block includes power chain for bias distribution and a data acquisition processing chain. The main Power Supply provides the low voltages to the detector electronics and the high bias voltages for PMTs and silicon modules. 

All the electronics is designed with embedded \virgolette{Hot/Cold} redundancy and all the components of the board have been selected capable to withstand a \SI{-40}{\celsius} to \SI{85}{\celsius} operating range. The maximum data transfer rate from the satellite is 50 GB per day. 

\begin{figure}[htb] 
\centering 
\includegraphics[width=\linewidth]{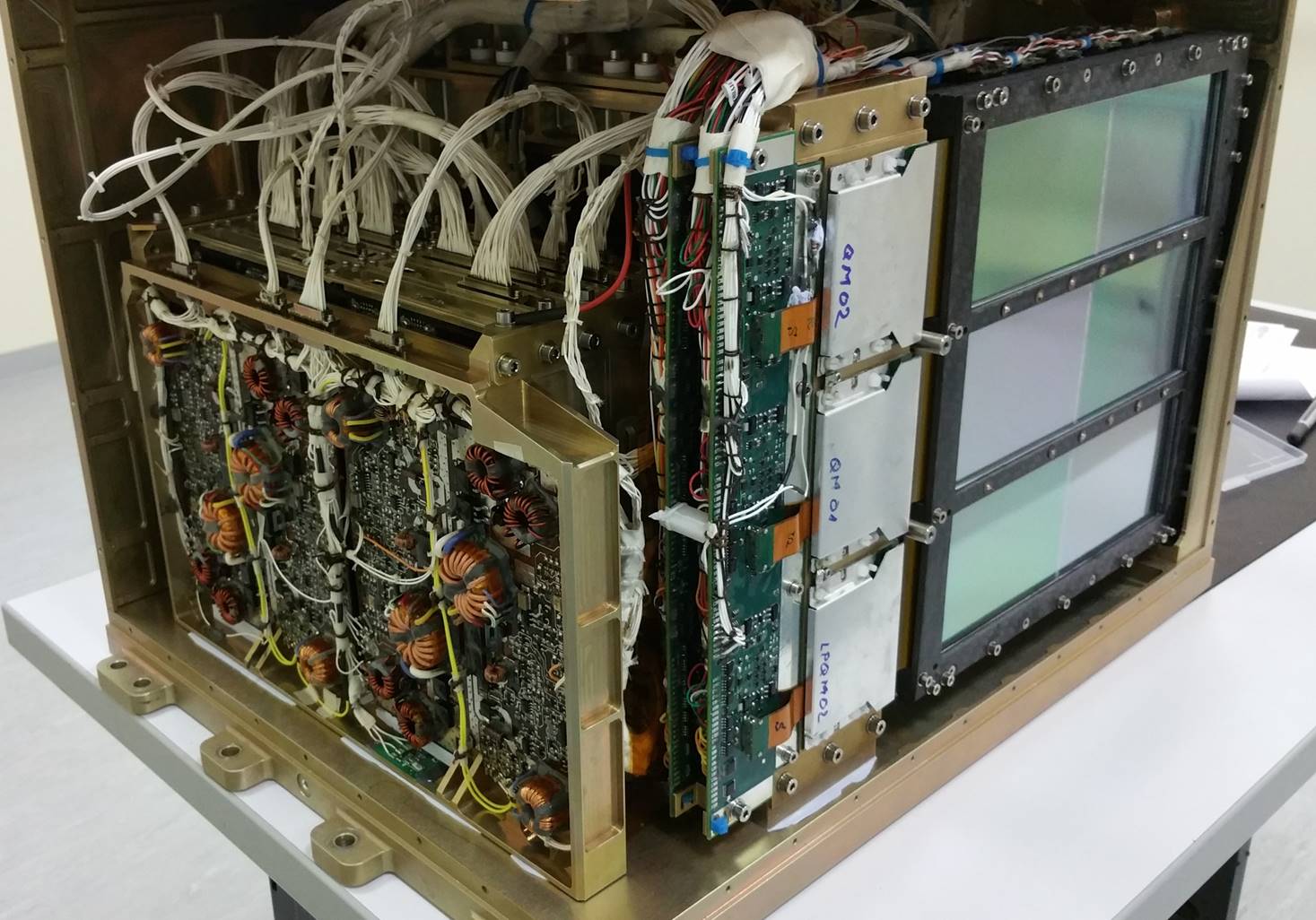}
\caption{The electronics and the silicon detector of the HEPD. }
\label{fig:electronics}. 
\end{figure}

The whole electronics system is schematized in figure \ref{fig:block}. It is composed by front-end electronics and four main boards: 
\begin{figure}[htb] 
\centering 
\includegraphics[width=\linewidth]{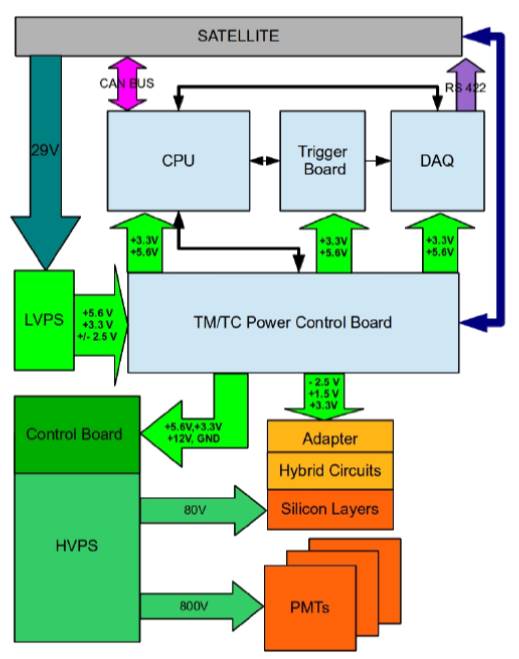}
\caption{A block diagram of the HEPD electronics subsystem. }
\label{fig:block}. 
\end{figure}
\begin{itemize}
\item Data Acquisition (DAQ) Board: manages all the scientific data of HEPD. The DAQ accomplies the following functionalities: acquisition of trigger signal from PMT/Trigger board, management of hybrid circuits on the silicon planes, acquisition of silicon planes data, computing of PMTs data and silicon planes data, data compression, transmission of scientific data on the scientific data link.
\item Trigger Board: manages the analog signals coming from the PMTs and generates the trigger signals needed for data acquisition. The main functions of this board are to invert and attenuate the PMTs analog signal to adapt it to input requirements of the EASIROC Integrated Circuits, to convert the EASIROC readout signals into digital signals, to allow the DAQ board to read the EASIROC digital output, to allow the CPU to configure the EASIROC, to generate and transmit \virgolette{slow} event trigger signals manipulating the \virgolette{fast} trigger signals coming from EASIROC, to allow the CPU to configure the \virgolette{slow} trigger generation algorithm.
\item CPU Board: controls the detector and communicates with the platform of the satellite via CAN BUS interface. The CPU manages the following functionalities: communication with Satellite OBDH computer via CAN bus, storage of non volatile information, management -via internal \virgolette{slow} control links bus- of TM/TC and LVPS control board, Trigger Board and DAQ Board, management of system diagnostic routines and of system configuration, system monitor.
 \item Telemetry/Telecommand board.

\end{itemize} 
 
\begin{figure}[htb] 
\centering 
\includegraphics[width=\linewidth]{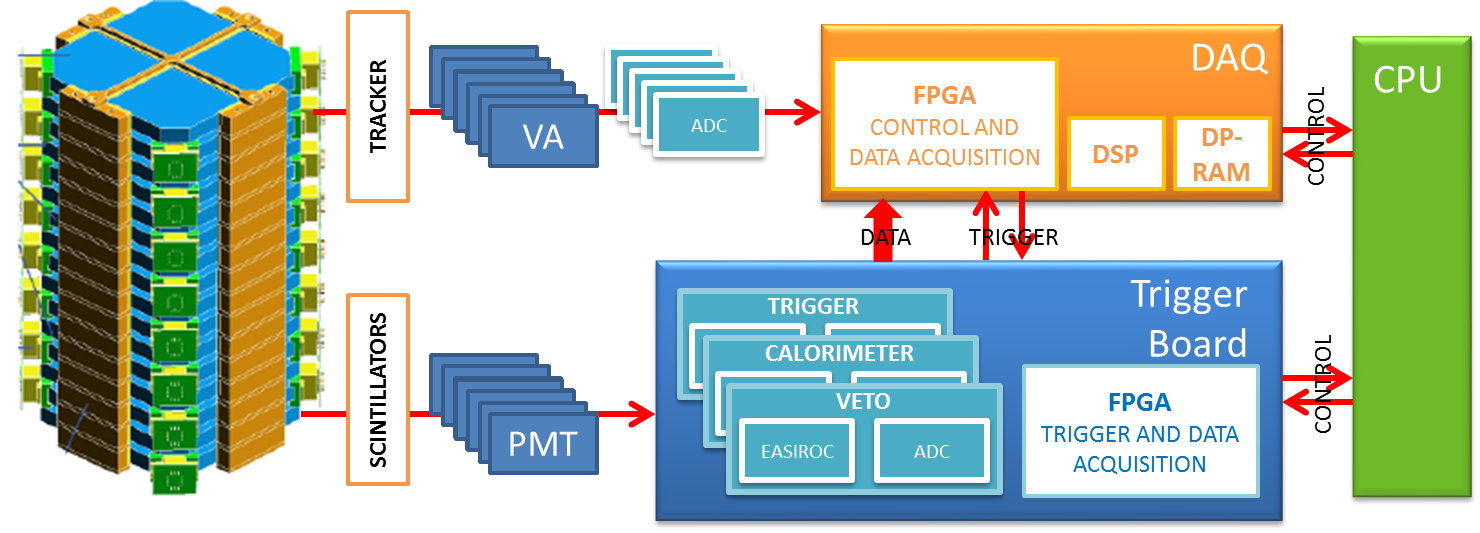}
\caption{A scheme of the data acquisition procedure. }
\label{fig:cses_electronics}. 
\end{figure}

A scheme of the data acquisition procedure is depicted in figure \ref{fig:cses_electronics}. The analogical signal read out from the PMTs associated to scintillator detectors are transmitted directly to the Trigger Board. Signals of each data processing block related to scintillators are managed by an FPGA which issues the FAST trigger signal needed to start the acquisition of the tracker by DAQ. After an handshake protocol, if the trigger is confirmed by DAQ, the Trigger Board sends Scintillators data to DAQ Board. Scintillators and tracker data are processed by a dedicated DSP and the results are written on a DP-RAM waiting to be transferred to satellite via RS-422 on a CPU command. 

\section{Test and qualification campaign}

\begin{figure}[htb] 
\centering 
\includegraphics[width=0.9\linewidth]{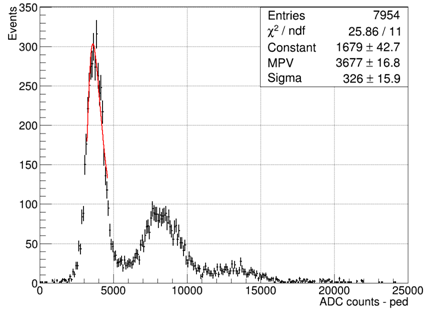}
\caption{Total energy loss in the HEPD QM plastic scintillator calorimeter for 30 MeV electrons. The red curve is a Landau fit to the one-particle peak of the distribution. }
\label{fig:e30mev}. 
\end{figure}

According to Chinese space procedures, the HEPD project involved the construction of four detector versions: the Electrical Model, the Structural and Thermal Model, the Qualification Model and the Flight Model. All the models have been built, assembled and integrated. In Spring 2016 we started the test and qualification campaign with the HEPD QM: vibration test at SERMS laboratory in Terni (PG) simulating launch and flight, thermal and vacuum test at SERMS laboratory simulating space environment. Finally, beam test were carried out at Beam Test Facility of the \virgolette{Laboratori Nazionali di Frascati} of INFN. The detector was irradiated with electrons and positrons from 30 to 150 MeV. The objective was to study the instrument response to electrons in the energy range of interest and to perform precise calibration of the calorimeter energy measurement for the QM. In Figure \ref{fig:e30mev} we show a preliminary plot of the total energy loss in the HEPD QM plastic scintillator calorimeter for 30 MeV electrons. The red curve is a Landau fit to the one-particle peak of the distribution.

The same tests have been taken with the HEPD FM. Beam test data are under analysis.

\section{Conclusions}

The HEPD has been developed by the Italian CSES-Limadou Collaboration. In this paper the main features of the HEPD have been described, with a scpecific focus on the electronics of the instrument. At the moment the FM of the HEPD is in China for pre-flight test.

The launch campaign will start in July 2017. The memorandum between Italy and China foresees also the commissioning post launch at CSES Ground Segment in Beijing and beam test of the QM after the redelivery.

\bigskip 

\begin{thebibliography}{9}   

\bibitem{Wang} Wang L. et al., Earthq Sci (2015) 28 4, 303
\bibitem{Zhang} Zhang X. et al., Nat. Hazards Earth Syst. Sci. (2013) 13, 197 
\bibitem{Sgrigna} Sgrigna V. et al., Journal of Atmospheric and Solar-Terrestrial Physics (2005), 67 1448
\bibitem{pamela} Adriani, O. et al., Physical Review Letters (2013), DOI: 10.1103/PhysRevLett.111.081102



\end{thebibliography}

\end{document}